\documentclass{article}
\usepackage{jheppub}
\usepackage{dcolumn}
\usepackage{bm}
\usepackage{verbatim}

\title{Study of Non-minimal SUSY SU(5) Model with Realistic Fermion Sectors}
\author{Yunfei Wu}
\emailAdd{yunfei\_wu@pku.edu.cn}
\author{and Da-Xin Zhang}
 \emailAdd{dxzhang@phy.pku.edu.cn}
\affiliation{School of Physics and State Key Laboratory of Nuclear Physics and Technology, Peking University,  Beijing 100871, China\\}
\abstract{We study a supersymmetric SU(5) model with the extra Higgs multiplets of $45+\overline{45}$. The unification of the gauge couplings, the fermion masses and the proton lifetime are discussed in details. The dimension-five operators mediated by different colored Higgs sector can be destructive with each other. This effect serves a way of solving the longevity of proton. We analytically analyze this destructive effect in a special limit where the mixing between the 5- and 45-plets is small. Although the theory does not hold in this special limit, it is a revelatory starting point. We can relax this limit and retain the destructive effect. In a generalized parameter space, this model is in accord with experimental results.
}

\begin{document}
\maketitle
\section{Introduction}

A successful model of grand unified theory(GUT) needs to satisfy at least the following key points: unification of the gauge couplings, realistic fermion masses and mixing, and long-lived proton. The first point of unification of the gauge couplings is realized in the class of models of supersymmetric(SUSY) GUT models\cite{luo,Ellis1991131}. To fulfill the second point, the Higgs sector needs extending beyond the minimal SU(5) model. For example, in the model by Georgi-Jarlskog\cite{Georgi:1979df}, 45-plet is added to generate the correct texture of fermion masses. Third, since in the SUSY GUT models the proton lifetime is determined by the dimension-five operators\cite{sakai,Weinberg} mediated by the colored Higgs multiplets, the Yukawa couplings need adjusting to suppress proton decay.

In this work we will study proton lifetime and fermion masses in a SUSY SU(5) model following Georgi-Jarlskog\cite{Georgi:1979df}. Originally, it was aiming at the fermion masses. Its SUSY version\cite{susygj,Raby:1992vk} contains a pair of extra Higgs multiplets of $45+\overline{45}$. After the SU(5) group is broken, the $45+\overline{45}$ contains a pair of weak doublets and two pairs of color triplets. This pair of doublets mixes with those from $5+\bar 5$.
%By a special choice of parameters, only one pair of weak doublets are light,
%so that the low energy limit of this model is the Minimal SUSY Standard Model(MSSM).
In order to make sure that the low energy limit of this model is the Minimal SUSY Standard Model(MSSM), we set a restriction on some parameters to let only one pair of weak doublets light. This model can give realistic texture of fermion masses. So we can assign the experimental constraints on the parameters related to fermion masses. The authors of Ref.\cite{Muraya3} claim that present proton decay experiments exclude the Minimal SUSY SU(5) GUT Model(MSGUT), although to some extent it still can be reconcilable with observation\cite{bajc}. We are trying to find some way to ensure longevity of proton.  By adding the new pair of $45+\overline{45}$, the contributions from different colored Higgs can be destructive. This effect have been introduced in Ref\cite{PhysRevD.77.015015}, and we will make a much detailed analysis here. This destructive effect can be analytically and easily shown in a special limit, called as "small mixing limit" here, where the mixing between 5 and 45-plets is small. By relaxing this limit, we can retain the destructive effects to ensure long enough proton lifetime.

This paper is organized as follows: The Higgs contents and their masses are given in Sec. \ref{sec:higgs}.  We set restricts on the heavy fields by requiring the gauge coupling unification in Sec. \ref{sec:gutm}. The Yukawa couplings with the matter fields are listed in Sec. \ref{sec:yuka}. The dimension-five operators are also given in this section.  In Sec. \ref{sec:gutc}, we formally analyze the proton decay width in  small mixing limit. In Sec. \ref{sec:num}, we give numerical results both in and beyond the small mixing limit. This model is in accord with experiments. Finally, we summarize our results.

\section{\label{sec:higgs}The Higgs contents }

In this section, we analyze the Higgs sectors of the model both before and after the breaking of SU(5) gauge symmetry. Before the symmetry breaking, the model have $5+\bar5$ and $45+\overline{45}$ Higgs multiplets, which is just the SUSY version of Georgi-Jarlskog Model\cite{Georgi:1979df}. After the breaking, the model becomes the MSSM at low energy.
\subsection{General Superpotential}
The general renormalizable  superpotential for the Higgs sector is
\begin{eqnarray}\label{WHiggs}
W_{Higgs}&=&\frac{1}{3}f\textnormal{Tr}\Sigma^3+\frac{1}{2}fV\textnormal{Tr}\Sigma^2+\lambda\bar
5_\alpha(\Sigma^\alpha_\beta+3V\delta^\alpha_\beta)5^\beta+\mu_1\bar 5_\alpha 5^\alpha\nonumber\\
&+&\rho
45^{\beta\gamma}_\alpha\{[\Sigma^\alpha_\rho\delta^{\sigma\delta}_{\beta\gamma}+\delta^\alpha_\rho(\Sigma^\sigma_\beta\delta^\delta_\gamma-\Sigma^\delta_\beta\delta^\sigma_\gamma)]+3V\delta^\alpha_\rho\delta^{\sigma\delta}_{\beta\gamma}\}\overline{45}^\rho_{\sigma\delta}+\frac{\mu_2}{2}45^{\beta\gamma}_\alpha\delta^{\sigma\delta}_{\beta\gamma}\overline{45}^\rho_{\sigma\delta}\nonumber\\
&+&\frac{1}{2V}\kappa_1 5^{[\alpha}\Sigma_\gamma^{\beta]}\overline{45}_{\alpha\beta}^\gamma+\frac{1}{2V}\kappa_2\bar 5_{[\alpha}\Sigma^\gamma_{\beta]}45^{\alpha\beta}_\gamma,
\end{eqnarray}
where $\Sigma$ is the adjoint Higgs. The Greek letters run from 1
to 5. The square brackets denote symmetrization. The third line corresponds the mixing between 5 and 45
Higgses. $\delta^{\sigma\delta}_{\beta\rho}$ is the extended Kronecker's delta defined as
$\delta^{\sigma\delta}_{\beta\rho}=\delta^\sigma_{\beta}
{\delta^\delta_\rho} - \delta^\sigma_\rho
{\delta^\delta_\beta}$.   Parts of this superpotential have already been listed in Refs.\cite{nonmini,nonm3,adj,adj2}.

The fields contents of the Higgs in $\overline {45}$  are
\begin{eqnarray}
\overline{45}&\rightarrow&\frac{\sqrt{3}}{2\sqrt{2}}\bar
H_a^\prime(2,1)_{-\frac{1}{2}}+\frac{\sqrt{2}}{2}\bar
H_\alpha^\prime(1,\bar 3)_{\frac{1}{3}}+\frac{\sqrt{2}}{2}H^a_{b\alpha}(3,\bar 3)_\frac{1}{3}+\frac{\sqrt{2}}{2}H^{\prime\prime\alpha}
\epsilon_{ab}(1,3)_{-\frac{4}{3}}\nonumber\\
&&+\frac{1}{2}H^{a\gamma}\epsilon_{\alpha\beta\gamma}(2,3)_\frac{7}{6}+\frac{1}{2}H^{\gamma\delta}_{(s)}\epsilon_{\alpha\beta\delta}(1,6)_\frac{1}{3}+\frac{\sqrt{2}}{2}H^\beta_{\alpha
a}(2,8)_{-\frac{1}{2}}.
\end{eqnarray}
All the coefficients are the normalization factors. %\cite{Babu1}.
The numbers in the brackets are the SU(2) and SU(3) representation dimensions and the subscripts are the U(1) charges. $s$ in round brackets denote symmetrization of the Greek indices. On the right-hand side of the above equation, the Latin letters take 1 and 2, while  the Greek letters take 1,2, and 3.
The Higgs contents in 45 have the corresponding conjugate terms.

The adjoint Higgs $\Sigma$ acquires a VEV diag$(2,2,2,-3,-3)V$ and breaks  SU(5) to the Standard Model(SM) gauge group\cite{Muraya}. The bilinear terms of 5 and 45 Higgs constituents in the SM gauge groups are
\begin{eqnarray}\label{mass}
W_{Higgs}^{mass}&=&(5\lambda V+\mu_1) \bar H_\alpha H^\alpha+ (\frac{13}{2}\rho V+\mu_2) \bar
H_\alpha^\prime H^{\prime\alpha}+(2\rho V+\mu_2) H^a_{b\alpha}H^{b\alpha}_a\nonumber\\
&&+(7\rho V+\mu_2)
H_\alpha^{\prime\prime} H^{\prime\prime\alpha}+(5\rho V+\mu_2) H_{a\gamma}H^{a\gamma}+(10\rho V+\mu_2)
H^{\gamma\delta}_{(s)}{H_{\gamma\delta}^{(s)}}\nonumber\\
&&+(7\rho V+\mu_2) H^\alpha_{\beta a}H_\alpha^{\beta a}+\mu_1
H^a\bar H_a+\mu_2  H^{\prime a}\bar H^\prime_a,
\end{eqnarray}
and the mixing terms between 5 and 45 Higgs are
\begin{eqnarray}\label{mix}
W_{Mixing}&=&\frac{\sqrt{2}}{2}\kappa_1 H^\alpha \bar
H^{\prime}_{\alpha}+\frac{\sqrt{3}}{2\sqrt{2}}\kappa_1 H^a\bar H_a^\prime\nonumber\\
&&+\frac{\sqrt{2}}{2}\kappa_2 \bar H_\alpha
H^{\prime\alpha}+\frac{\sqrt{3}}{2\sqrt{2}}\kappa_2 \bar H_a H^{\prime a}.
\end{eqnarray}

From Eq.({\ref{mass}}), we can see that this model has two pairs of weak doublets $(H^a,\bar H_a)$ and $(H^{\prime a},\bar H^{\prime}_a)$ and three pairs of colored Higgs triplets $(H^\alpha,\bar H_\alpha)$, $(H^{\prime\alpha},\bar H_\alpha^\prime)$, and $(H^{\prime\prime}_\alpha,H^{\prime\prime\alpha})$. The unprimed ones come from $5+\bar 5$ Higgs multiplets and the primed ones come from $45+\overline{45}$ Higgs multiplets. Mixing between primed and unprimed ones take place when $\kappa$'s are nonzero. Note that  $(H^{\prime\prime}_\alpha,H^{\prime\prime\alpha})$ do not mix with other Higgs triplets. The masses of these sectors are exhibited in the following context.

\subsection{Doublet Masses}
The Higgs
doublets' mass and mixing terms can be achieved from Eqs.(\ref{mass}) and (\ref{mix}) and can be rewritten as 
\begin{equation}
M_D=\mu_1
H^a\bar H_a+\mu_2  H^{\prime a}\bar H^\prime_a+\frac{\sqrt{3}}{2\sqrt{2}}\kappa_1 H^a\bar H_a^\prime+\frac{\sqrt{3}}{2\sqrt{2}}\kappa_2 \bar H_a H^{\prime a}.
\end{equation}
The mass matrix can be diagonalized with the following rotation
\begin{eqnarray}\label{rotatedhiggs}
 \left[\begin{array}{c}(H_{u},H_d)\\ (H^{\prime}_{u},H^{\prime}_d)
\end{array}\right]& =&\left[\begin{array}{cc} \cos\theta_{
D}& \sin\theta_{D}\\ -\sin\theta_{D}&\cos\theta_{
D}\end{array}\right]\left[\begin{array}{c}(H^{a}, \bar H_{a})\\ (H^{\prime
a}, \bar H_{a}^\prime)
 \end{array}\right].
\end{eqnarray}
The fields on the left-hand side of Eq.({\ref{rotatedhiggs}})
represents the mass eigenstates.
If we set
\begin{equation}\label{mcond}
3\kappa_1\kappa_2=8\mu_1\mu_2,
\end{equation}
we get one pair of Higgs doublet $(H_u,H_d)$ to be massless and the other pair $(H'_u,H'_d)$ has squared mass eigenvalue
\begin{equation}\label{mhd}
m_{H_D'}^2={(\mu_1^2+\mu_2^2)+\frac{3}{8}(\kappa_1^2+\kappa_2^2)}.
\end{equation}
It must be emphasized that the condition of Eq.(\ref{mcond}) is still a typical fine tuning of parameters.
Under this condition, the doublets' rotating angle can be written as
\begin{eqnarray}
\tan\theta_D=\frac{\sqrt{6}\kappa_2}{4\mu_2}.
\end{eqnarray}
%The condition Eq.(\ref{mcond}) realizes the doublet-triplet mass splitting. This condition is basic postulate in the rest of this work.
When $\mu_1\ll\mu_2$, slight breaking of Eq.(\ref{mcond}) ensures a small eigenvalue of
\begin{equation}
\mu\approx\mu_1-\frac{3\kappa_1\kappa_2}{8\mu_2},
\end{equation}
which is just the coefficient of "$\mu$-term" in the MSSM.
%This is a possible solution to the smallness of $\mu$ through the see-saw mechanism.
Then at low energy there exist only one pair of Higgs doublets. They are identical to the Higgs doublets of the MSSM. The flavor changing neutral currents mediated by the heavy Higgs are then negligible. The condition Eq.(\ref{mcond}) realizes the doublet-triplet mass splitting. This condition is basic postulate in the rest of this work.

\subsection{Triplet Masses}
 This model have three pair of colored triplets. For $(H^{\prime\prime}_\alpha,H^{\prime\prime\alpha})$ do not mix with the others, their masses are just the bilinear term in Eq.(\ref{mass}). We only have to consider the other two pairs' mixing. They can be written as 
\begin{equation}
M_T=(5\lambda V+\mu_1) \bar H_\alpha H^\alpha+ (\frac{13}{2}\rho V+\mu_2) \bar
H_\alpha^\prime H^{\prime\alpha}+\frac{\sqrt{2}}{2}\kappa_1 H^\alpha \bar
H^{\prime}_{\alpha}+\frac{\sqrt{2}}{2}\kappa_2 \bar H_\alpha
H^{\prime\alpha}.
\end{equation}
The mass matrix of Higgs triplets can be diagonalized with the following rotation
\begin{eqnarray}\label{rotatedhiggs2}
 \left[\begin{array}{c}(H_{c},\bar H_c )\\ (H_{c}^\prime,\bar H_c^\prime)
\end{array}\right]& =& \left[\begin{array}{cc}\cos\theta_{T}&\sin\theta_{T}\\-\sin\theta_{T}&\cos\theta_{
T}\end{array}\right]\left[\begin{array}{c}(H^{\alpha},\bar H_{\alpha})\\
(H^{\prime\alpha}, \bar H_{\alpha}^\prime)
 \end{array}\right].
\end{eqnarray}
The mass eigenvalues of the rotated Higgs triplets are
\begin{eqnarray}\label{mhc}
M_{H_c}=\big(\sqrt{2(\kappa_1-\kappa_2)^2+s^2}+\sqrt{2(\kappa_1+\kappa_2)^2+t^2}\big)/4,\nonumber\\
M_{H_c'}=\big|\sqrt{2(\kappa_1-\kappa_2)^2+s^2}-\sqrt{2(\kappa_1+\kappa_2)^2+t^2}\big|/4,
\end{eqnarray}
where
\begin{eqnarray}
s=13 \rho  V + 10 \lambda V  +2 \mu _2+2 \mu _1,\nonumber\\
t=13 \rho  V - 10 \lambda V  +2 \mu _2-2 \mu _1.
\end{eqnarray}
We can express the rotating angle as
\begin{eqnarray}
\tan\theta_T=\frac{\sqrt{[2(\kappa_1+\kappa_2)^2+t^2][2(\kappa_1-\kappa_2)^2+s^2]}-st-2(\kappa_1^2-\kappa_2^2)}{\sqrt{2}[(\kappa_1+\kappa_2)s-(\kappa_1-\kappa_2)t ]}.
\end{eqnarray}

 At the end of this section, we summarize all the heavy Higgs masses in Table {\ref{tab:sum}}. These masses enter into the running the gauge couplings and the colored ones can mediate nucleon decay.

\begin{table}[h]
\caption{\label{tab:sum} Summary of the GUT-scale Higgs masses in terms of basic parameters.}
\begin{tabular}{llc}
\hline\hline
Higgs&Representation&\\
multiplets &under SM group&\raisebox{1.5ex}[0pt]{Mass} \\
\hline
${H_c}\,\bar H_c$&$(1,3)_{-\frac{1}{3}}\,(1,\bar3)_\frac{1}{3}$&$\big(\sqrt{2(\kappa_1-\kappa_2)^2+s^2}+\sqrt{2(\kappa_1+\kappa_2)^2+t^2}\big)/4$\\
${H_c'}\,{\bar H_c'}$&$(1,3)_{-\frac{1}{3}}\,(1,\bar3)_\frac{1}{3}$&$\big|\sqrt{2(\kappa_1-\kappa_2)^2+s^2}-\sqrt{2(\kappa_1+\kappa_2)^2+t^2}\big|/4$\\
${H_c''\, \bar H_c''}$&$(1,3)_{-\frac{4}{3}}\,(1,\bar3)_{\frac{4}{3}}$&$7\rho V+\mu_2$\\
${H^\alpha_{\beta a}}\, H_\alpha^{\beta a}$&$(2,8)_{-\frac{1}{2}}\,(2,8)_{\frac{1}{2}}$&$7\rho V+\mu_2$\\
${H^a_{b\alpha}}\,H_a^{b\alpha}$&$(3,\bar3)_{\frac{1}{3}}\,(3,3)_{-\frac{1}{3}}$&$2\rho V+\mu_2$\\
${H^{\alpha\beta}_{(s)}}\,H_{\alpha\beta}^{(s)}$&$(1,6)_{\frac{1}{3}}\,(1,\bar 6)_{-\frac{1}{3}}$&$10\rho V+\mu_2$\\
${H^{a\beta}}\,H_{a\beta}$&$(2,3)_{\frac{7}{6}}\,(2,\bar 3)_{-\frac{7}{6}}$&$5\rho V+\mu_2$\\
${H_u'}\,H_d'$&$(2,1)_{\frac{1}{2}}\,(2,1)_{-\frac{1}{2}}$&$\sqrt{(\mu_1^2+\mu_2^2)+\frac{3}{8}(\kappa_1^2+\kappa_2^2)}$\\
\hline\hline
\end{tabular}
\end{table}

\section{\label{sec:gutm}Constraints on GUT-scale Masses}

In this section, we examine the gauge coupling unification and we get limits on the GUT-scale masses from the requirements of the unification. The theoretical perturbative bounds on Yukawa couplings can also put further constraints on GUT-scale masses\cite{Muraya}.

The running of the three gauge coupling constants in MSGUT
have already been investigated\cite{Muraya2}. By adding the extra $45+\overline{45}$ Higgs
effects, we get the new formulae
 \begin{eqnarray}
\alpha_3^{-1} (m_Z) &=& \alpha_{5}^{-1} (\Lambda)
	+ \frac{1}{2\pi} \bigg\{
		\left( -2 - \frac{2}{3} N_g \right) \ln \frac{m_{SUSY}}{m_Z}
		+ (-9 + 2 N_g) \ln \frac{\Lambda}{m_Z}
			 \nonumber \\
%\phantom{\alpha_{5}^{-1} (\Lambda) + \frac{1}{2\pi} \{ }
&&-4 \ln \frac{\Lambda}{M_V} + 3 \ln \frac{\Lambda}{M_\Sigma}
		+ \ln \frac{\Lambda}{M_{H_c}}+\ln\frac{\Lambda}{M_{H_c'}}+\ln\frac{\Lambda}{M_{H_c''}}\nonumber\\
&&+3\ln\frac{\Lambda}{M_{H_{b\alpha}^a}}+2\ln\frac{\Lambda}{M_{H^{a\beta}}}+5\ln\frac{\Lambda}{M_{H^{\alpha\beta}_{(s)}}}+12\ln\frac{\Lambda}{M_{H_{\beta a}^\alpha}} \bigg\},
\end{eqnarray}
\begin{eqnarray}
\alpha_2^{-1} (m_Z) &=& \alpha_{5}^{-1} (\Lambda)
	+ \frac{1}{2\pi} \bigg\{
		\left( - \frac{2}{3} N_g - \frac{13}{6} \right)
			 \ln \frac{m_{SUSY}}{m_Z}
		+ (-5 + 2 N_g ) \ln \frac{\Lambda}{m_Z}
			 \nonumber \\
% \phantom{\alpha_{5}^{-2} (\Lambda) + \frac{1}{2\pi} \{ }
&&+\ln\frac{\Lambda}{m_{H_D'}}-6 \ln \frac{\Lambda}{M_V} + 2 \ln \frac{\Lambda}{M_\Sigma}+ 12 \ln \frac{\Lambda}{M_{H_{b\alpha}^a}}+3\ln \frac{\Lambda}{M_{H^{a\beta}}}+8\ln \frac{\Lambda}{M_{H^{\alpha}_{\beta a}}}\bigg\},
\end{eqnarray}
\begin{eqnarray}
\alpha_1^{-1} (m_Z) &=& \alpha_{5}^{-1} (\Lambda)
	+ \frac{1}{2\pi} \bigg\{
		\left( -\frac{2}{3} N_g - \frac{1}{2} \right)
			 \ln \frac{m_{SUSY}}{m_Z}
		+ \left(2 N_g + \frac{3}{5} \right)
			\ln \frac{\Lambda}{m_Z}
			 \nonumber \\
% \phantom{\alpha_{5}^{-3} (\Lambda) + \frac{1}{2\pi} \{ }
&&+\frac{3}{5}\ln\frac{\Lambda}{m_{H_D'}}-10 \ln \frac{\Lambda}{M_V}
		+ \frac{2}{5} \ln \frac{\Lambda}{M_{H_c}}+\frac{2}{5}\ln\frac{\Lambda}{M_{H_c'}}+\frac{32}{5}\ln\frac{\Lambda}{M_{H_c''}}\nonumber\\
&&+\frac{6}{5}\ln\frac{\Lambda}{M_{H_{b\alpha}^a}}+\frac{49}{5}\ln\frac{\Lambda}{M_{H^{a\beta}}}+\frac{4}{5}\ln\frac{\Lambda}{M_{H^{\alpha\beta}_{(s)}}}+\frac{24}{5}\ln\frac{\Lambda}{M_{H_{\beta a}^\alpha}}  \bigg\}.
\end{eqnarray}
Here, the scale $\Lambda$ is larger than any of the GUT-scale
masses and all the masses except for $m_{SUSY}$ and $m_Z$ are taken around the GUT-scale.
We can get all the GUT-scale masses' dependence on the parameters of superpotential from Eqs.(\ref{mass}), (\ref{mhd}), and (\ref{mhc}).
The number of generation $N_g$ is three. We take all MSSM
particles' masses at $m_{SUSY}$. By eliminating $\alpha_5^{-1}$, we have
\begin{eqnarray}\label{gut1}
(3\alpha_2^{-1}-2\alpha_3^{-1}-\alpha_1^{-1})(m_Z)&=&\frac{1}{2\pi}\Big\{-2\ln\frac{m_{SUSY}}{m_Z}-\frac{12}{5}\ln\frac{m_{H'_D}}{m_Z}\nonumber\\
&&+\frac{12}{5}\ln\frac{M_{H_c}M_{H_c'}}{m^2_Z}+\frac{6}{5}\ln\frac{M^4_{H^{a\beta}}M^9_{H^{\alpha\beta}_{(s)}}M^4_{H^\alpha_{\beta
    a}}M^7_{H_c''}}{M^{24}_{H^a_{b\alpha}}}\Big\},
\end{eqnarray}
\begin{eqnarray}\label{gut2}
(5\alpha_1^{-1}-3\alpha_2^{-1}-2\alpha_3^{-1})(m_Z)&=&\frac{1}{2\pi}\Big\{8\ln\frac{m_{SUSY}}{m_Z}+12\ln\frac{M_V^2M_\Sigma}{m_Z^3}\nonumber\\
&&+6\ln\frac{M^6_{H^a_{b\alpha}}M^4_{H^\alpha_{\beta
    a}}M_{H^{\alpha\beta}_{(s)}}}{M^6_{H^{a\beta}}M^5_{H_c''}}\Big\}.
\end{eqnarray}
The Eqs.(\ref{gut1},\ref{gut2}) give restrictions on the GUT-scale mass spectrum from the weak-scale parameters. But the restrictions here
are much looser than those in Refs.\cite{Barbieri,Ellis}. The mass splitting among the MSSM particles also affects the GUT-scale mass spectrum.
The detailed analysis of SUSY mass splitting can be found in Ref.\cite{Muraya}.
If $\rho V \ll \mu_2$, the masses of Higgs
multiplets from the 45-plet are highly degenerate except for $H_D'$ and
$H_c'$. When we restrict to the universal scalar mass dominates the
SUSY breaking and take two-loop gauge couplings into account, we have
the following constraints
\begin{equation}\label{rcond0}
1.7\times10^{16}\text{GeV}\le(M_V^2M_\Sigma)^{1/3}(1-2.5\frac{\rho V}{\mu_2})\le 2.0\times10^{16}\text{GeV},
\end{equation}
\begin{equation}\label{rcond1}
3.5\times10^{14}\text{GeV}\le
\frac{M_{H_c}M_{H_c'}}{m_{H_D'}}(1+69.5\frac{\rho V}{\mu_2})\le3.6\times 10^{15}\text{GeV},
\end{equation}
for gluino mass $100\text{GeV}\le m_{\tilde g}\le 1 \text{TeV}$. %If we further assume $\kappa_{1(2)}\ll\mu_{2}$ and $t\ll s$, we have
%\begin{equation}\label{M33}
%M_{H^a_{b\alpha}}<3.6\times10^{15}\text{GeV}.
%\end{equation}
 If $\rho V
\gg \mu_2$, the constraints become
\begin{equation}
4.0\times10^{16}\text{GeV}\le(M_V^2M_\Sigma)^{1/3}(1+0.29\frac{\mu_2}{\rho V})\le 4.7\times10^{16}\text{GeV},
\end{equation}
\begin{equation}\label{rcond2}
4.4\times10^{7}\text{GeV}\le
\frac{M_{H_c}M_{H_c'}}{m_{H_D'}}(1-4.36\frac{\mu_2}{\rho V})\le 4.2\times 10^{8}\text{GeV}.
\end{equation}
The above  equations are the extended version of those from Ref.\cite{Muraya3}.

We define $M_{GUT}=(M_V^2M_\Sigma)^{1/3}$ and it is constrained
as
\begin{equation}
1.7\times10^{16}\text{GeV}\le M_{GUT}\le 4.7\times10^{16}\text{GeV}.
\end{equation}

The applicability of perturbation requires that the dimensionless
couplings $f$, $\lambda$, $\rho$ in Eq.(\ref{WHiggs}) to
be small.
%The light Higgs doublet condition Eq.(\ref{mcond}) gives $\mu_1\mu_2$
%small too.
The perturbative bound of $f$ under Planck
scale in Ref.\cite{Muraya} gave $M_V>0.56M_\Sigma$. Numerical study
gives
\begin{equation}
 M_\Sigma<6.9\times10^{16}\text{GeV},
\end{equation} and
\begin{equation}
M_V>1.4\times10^{16}\text{GeV}.
\end{equation}
The above equation satisfies the bound on dimension-six
operator mediated $p\to\pi^0+e^+$ which requires $M_V>(2.57-3.23)\times
10^{15}$GeV\cite{Nath}.

\section{\label{sec:yuka}The Yukawa Couplings and The Dimension-Five Operators}

In this section, we present the realistic fermion mass relations and the dimension-five operators. There are several Higgs multiplets which give contributions to the fermion masses or to the dimension-five operators, and these Higgs multiplets contribute in different ways.
%The fermion mass generation through the Georgi-Jarlskog mechanism only needs two pairs of Higgs doublets while the dimension-five operators have three pairs of color-Higgs multiplets' contributions.
So the correlation between the dimension-five operator couplings and the matter fields' Yukawa couplings is diluted.
\subsection{Superpotential}
%The Georgi-Jarlskog model provides a mechanism of generating realistic lepton-quark mass relations and the mixing angles.
Before the GUT gauge group symmetry is broken, the superpotential for the Yukawa couplings is
\begin{eqnarray}\label{Yukawa}
W&=&\sqrt{2}f^{ij}\psi^{\alpha\beta}_i\psi_{j\alpha}\bar
5_\beta+\frac{1}{4}h^{ij}\epsilon_{\alpha\beta\gamma\delta\epsilon}\psi^{\alpha\beta}_i\psi^{\gamma\delta}_j5^\epsilon\nonumber\\
&&+\sqrt{2}f^{\prime
  ij}\psi^{\alpha\beta}_i\psi_{j\gamma}\overline{45}^\gamma_{\alpha\beta}+2h^{\prime
  ij}\epsilon_{\alpha\beta\gamma\rho\sigma}\psi^{\alpha\beta}_i\psi^{\gamma\delta}_j45^{\rho\sigma}_{\delta},
\end{eqnarray}
where $\psi^{\alpha\beta}_i$ and $\psi_{j\alpha}$ are the 10- and $\bar 5$-plet
matter fields\cite{Georgi,Muraya} with $i$ and $j$ as the generation
indices. Here the Greek letters run from 1 to 5. We work in the basis that the mass matrix for the up-type quarks are already diagonalized to reduce complexity.

The coupling constants are defined as
\begin{eqnarray}
h^{(\prime) ij}&=&h^{(\prime)i}e^{i\phi_i}\delta^{ij},\nonumber\\
f^{(\prime)ij}&=&V_{ij}^*f^{(\prime)j},
\end{eqnarray}
where $V_{ij}$'s are the Cabibbo-Kobayashi-Maskawa(CKM)
matrix elements.
The phases $\phi_i$'s and the $V_{ij}$'s are valid for both couplings to 5 and 45 Higgs, for they are fixed by the definition of the matter fields. Even if we set different $\phi_i'$'s and $V_{ij}'$'s for the primed Yukawa couplings, their difference with unprimed ones can be absorbed into the redefinitions of the primed Yukawa couplings. Only two of the phase are independent and we follow \cite{Muraya} to take
\begin{equation}
\phi_u+\phi_c+\phi_t=0.
\end{equation}

Under the SM group, the superpotential for the matter fields can be deduced from Eq.(\ref{Yukawa}) and written as
\begin{eqnarray}\label{yuka}
W&=&(f^i\bar H_a+\frac{\sqrt{3}}{2\sqrt{2}}f^{\prime
  i}\bar H^\prime_a)e_i^cL_i\nonumber\\
&&+V^*_{ij}(f^j\bar H_a-\frac{\sqrt{3}}{6\sqrt{2}}f^{\prime j}\bar H^\prime_a)Q_id^c_j\nonumber\\
&&+(h^iH^a-\frac{\sqrt{6}}{2}h^{\prime i}H^{\prime a})u_i^cQ_i\nonumber\\
&&+(\frac{1}{2}h^{i}H^\alpha+2\sqrt{2}h^{\prime i}H^{\prime\alpha})e^{i\phi_i}Q_iQ_i\nonumber\\%+2\sqrt{2}h^{\prime i}e^{i\phi_i}Q^{c\alpha}_iQ^{a\beta}_iH^{b\gamma}_a\epsilon_{\alpha\beta\gamma}\epsilon_{bc}\nonumber\\
&&+V^*_{ij}(f^j\bar H_\alpha+\frac{\sqrt{2}}{2}f^{\prime j}\bar H^\prime_\alpha)Q_iL_j+\frac{\sqrt{2}}{2}V_{ij}^*f^{\prime j}Q^{b\alpha}_iL_{ja}H^a_{b\alpha}\nonumber\\
&&+\left[h^iH^\alpha+4\sqrt{2}h^{\prime i}H^{\prime\alpha}\right]V_{ij}u^c_ie^c_j\nonumber\\
&&+e^{-i\phi_i}V_{ij}^*(f^j\bar H_\alpha+\frac{\sqrt{2}}{2}f^{\prime j}\bar H^\prime_\alpha)u^c_id^c_j\nonumber\\
&&+\sqrt{2}f^{\prime i}e^c_id^c_{i \alpha}H^{\prime\prime\alpha}\nonumber\\
%-2\sqrt{2}h^{\prime i}e^{-i\phi_i}u^c_{i\alpha} u^c_{i\beta}H_\gamma^{\prime\prime}\epsilon^{\alpha\beta\gamma}
&&+e^{-i\phi_i}V_{ij}^*f^{\prime
  j}u^c_{i\gamma}L_{ja}H^{a\gamma}-4h^{\prime i}e^{i\phi_i}e^c_iQ^{a\gamma}_iH_{a\gamma}\nonumber\\
&&+e^{-i\phi_i}V_{ij}^*f^{\prime
  j}u^c_{i\gamma}d^c_{j\beta}H^{\beta\gamma}_{(s)}-2\epsilon_{ab}h^{\prime
  i}e^{i\phi_i}Q^{a\beta}_iQ^{b\gamma}_iH_{\beta\gamma}^{(s)}\nonumber\\
&&-\sqrt{2}V^*_{ij}f^{\prime j}Q_i^{a\alpha}d^c_{j\beta}H^\beta_{\alpha
  a}+2\sqrt{2}\epsilon_{ba}h^{\prime i}u^c_{i\alpha}Q^{b\beta}_iH^{\alpha a}_\beta,
\end{eqnarray}
where the Latin letters represent the SU(2) indices and the Greek letters represent the SU(3) indices. The coupling $\epsilon_{\alpha\beta\gamma}Q^{\alpha}  Q^{\beta} H^{b\gamma}_a$ vanishes for  $Q^\alpha Q^\beta$  symmetric under SU(2) transformation and anti-symmetric under SU(3) transformation. So $(3,3)$ and $(3,\bar3)$ components of $45+\overline {45}$ do not couple to $QQ$.

\subsection{Fermion Mass Texture}

The Georgi-Jarlskog model provides a mechanism of generating realistic lepton-quark mass relations.
The fermion mass formulae can be easily gotten from Eq.(\ref{yuka})
\begin{eqnarray}\label{fmass}
m_{u_i}&=&h^i\langle H^a\rangle-\frac{\sqrt{6}}{2}h^{\prime i}\langle
H^{\prime a}\rangle,\nonumber\\
m_{d_i}&=&f^i\langle \bar H_a\rangle-\frac{\sqrt{6}}{12}f^{\prime i}\langle
\bar H_a^\prime\rangle,\nonumber\\
m_{e_i}&=&f^i\langle \bar H_a\rangle+\frac{\sqrt{6}}{4}f^{\prime i}\langle
\bar H_a^\prime\rangle.
\end{eqnarray}
Together with Eq.(\ref{rotatedhiggs}), we get the Yukawa couplings of the light Higgs doublets to the light fermions as follows
\begin{eqnarray}\label{fyuka}
h_{u_i}&=&h^i\cos\theta_D-\frac{\sqrt{6}}{2}h^{\prime i}\sin\theta_D,\nonumber\\
f_{d_i}&=&f^i\cos\theta_D-\frac{\sqrt{6}}{12}f^{\prime i}\sin\theta_D,\nonumber\\
f_{e_i}&=&f^i\cos\theta_D+\frac{\sqrt{6}}{4}f^{\prime i}\sin\theta_D.
\end{eqnarray}

Here we can reexamine the reliability of one set of phases and CKM matrix elements. The masses of up-type quarks would be complex if $h^{\prime
  ij}$ have  different phases with $h^{ij}$. The masses of down-type quarks would be complex with $V^{\prime}_{ij}$'s different from CKM. So are the leptons. From above equations, we can see that $f^{(\prime)i}$ are fixed up to $\theta_D$ and $\tan\beta$, where $\tan\beta=\langle H_u\rangle/\langle H_d\rangle$. $h^{(\prime) i}$ have more degrees of freedom.

\subsection{Dimension-Five Operators and Effective Lagrangian}
We can get the dimension-five
operators that violate the baryon-lepton numbers, by integrating out the colored Higgs multiplets.
These operators can be put into two categories: the  LLLL-types ones by integrating out $H_c$ and $H_c'$  and the RRRR-type ones. The LLLL-type operators can be written explicitly as
\begin{equation}
W_{5L}=Y^{ij}e^{i\phi_i}V^*_{kj}(Q_iQ_i)(Q_kL_j),
\end{equation}
where
\begin{eqnarray}
Y^{ij}&=&\frac{1}{M_{H_c}}\Big(\frac{h^if^j}{2}\cos^2\theta_T+\frac{\sqrt{2}}{4}h^if^{\prime j}\cos\theta_T\sin\theta_T\nonumber\\
&&+2h^{\prime i}f^{\prime j}\sin^2\theta_T+2\sqrt{2}h^{\prime i}f^j\sin\theta_T\cos\theta_T\Big)\nonumber\\
&&+\frac{1}{M_{H'_c}}\Big(\frac{h^if^j}{2}\sin^2\theta_T-\frac{\sqrt{2}}{4}h^if^{\prime j}\cos\theta_T\sin\theta_T\nonumber\\
&&+2h^{\prime i}f^{\prime j}\cos^2\theta_T-2\sqrt{2}h^{\prime i}f^j\sin\theta_T\cos\theta_T\Big).
\end{eqnarray}
%The extra LLLL-type operators can be written as
%\begin{equation}
%W_{5L_2}=\frac{2}{M_{H^a_{b\alpha}}}h^{\prime i}e^{i\phi_i}V^*_{kj}f^{\prime j}(Q_kQ_i)(Q_iL_j).
%\end{equation}
The RRRR-type operators can be written as
\begin{equation}
W_{5_R}=X^{ij}V_{ij}e^{-i\phi_{k}}V^*_{kl}(u^c_ie^c_j)(u^c_kd^c_l),
\end{equation}
where
\begin{eqnarray}
X^{ij}&=&\frac{1}{M_{H_c}}\Big({h^if^j}\cos^2\theta_T+\frac{\sqrt{2}}{2}h^if^{\prime j}\cos\theta_T\sin\theta_T\nonumber\\
&&+4h^{\prime i}f^{\prime j}\sin^2\theta_T+4\sqrt{2}h^{\prime i}f^j\sin\theta_T\cos\theta_T\Big)\nonumber\\
&&+\frac{1}{M_{H'_c}}\Big(h^if^j\sin^2\theta_T-\frac{\sqrt{2}}{2}h^if^{\prime j}\cos\theta_T\sin\theta_T\nonumber\\
&&+4h^{\prime i}f^{\prime j}\cos^2\theta_T-4\sqrt{2}h^{\prime i}f^j\sin\theta_T\cos\theta_T\Big).
\end{eqnarray}
The trigonometric functions in $Y^{ij}$ and $X^{ij}$ hail from the  mass
matrix diagonalization.

The external sfermion legs of the dimension-five operators will be
converted to fermions at SUSY breaking scale, by dressing of gauginos
or doublet Higgsinos. We only focus on the LLLL-type operators here for simplicity, although the RRRR-type can also be important\cite{Goto}.  The dressing of wino to $(QQ)(QL)$ gives the
most important contribution\cite{Arno,Muraya}, and yields the triangle diagram factor
\begin{eqnarray}
f(u,d)&=&\frac{M_2}{m_{\tilde u}^2-m_{\tilde d}^2}\left(\frac{m_{\tilde u}^2}{m_{\tilde u}^2-M_2^2}\ln\frac{m_{\tilde u}^2}{M_2^2}\right.\nonumber\\
&&-\left.\frac{m_{\tilde d}^2}{m_{\tilde
d}^2-M_2^2}\ln\frac{m_{\tilde d}^2}{M_2^2}\right),\label{triangle}
\end{eqnarray}
where $M_2$ is the wino masses.
The resulting four-fermion operators are
\begin{eqnarray}\label{decay}
\mathcal L
&=&Y^{ik}\frac{\alpha_2}{\pi}e^{i\phi_i}V^*_{jk}\epsilon_{\alpha\beta\gamma}\nonumber\\
&&\times\left[(u^\alpha_id^{\prime\beta}_i)(d^{\prime\gamma}_j\nu_k)(f(u_j,e_k)+f(u_i,d^\prime_i))\right.\nonumber\\
&&+(d^{\prime\alpha}_i u^\beta_i)(u_j^\gamma e_k)(f(u_i,d^\prime_i)+f(d^\prime_j,\nu_k))\nonumber\\
&&+(d^{\prime\alpha}_i\nu_k)(d^{\prime\beta}_iu^\gamma_j)(f(u_i,e_k)+f(u_i,d^\prime_j))\nonumber\\
&&+\left.(u^\alpha_id^{\prime\beta}_j)(u^\gamma_ie_k)(f(d^\prime_i,u_j)+f(d^\prime_i,\nu_k))\right].
\end{eqnarray}
%where
%\begin{equation}\label{d5coupling}
%y^{ik}=\left(Y^{ik}-\frac{h^{\prime i}f^{\prime k}}{M_{H^a_{b\alpha}}}\right).
%\end{equation}
The total antisymmetry in color index requires $i\not=k$, which implies the dominant mode is $p\to K\bar{\nu}$\cite{Muraya}.
Taking
renormalization effects into account, the most relevant terms for $p\to K^++\bar\nu_\mu$ are
\begin{eqnarray}\label{fl}
\mathcal L&=&\frac{\alpha_2}{\pi}V^*_{us}A_L\epsilon_{\alpha\beta\gamma}\left((d^\alpha u^\beta) (s^\gamma \nu_\mu)+ (s^\alpha u^\beta)(d^\gamma \nu_\mu) \right)\nonumber\\
&&\times \left[A_S(c,u,s)e^{i\phi_c} Y^{22} V_{cs} V_{cd} (f(c,\,\mu) + f(c,\,d'))\right.\nonumber\\
&&\left.+ A_S(t,u,s)e^{i\phi_t} Y^{32} V_{ts} V_{td} (f(t,\,\mu) + f(t,\,d'))\right].
\end{eqnarray}
The function $A_S$ refers to the short range renormalization effect between the unification and the
SUSY breaking scales and $A_L$ the long range renormalization effect
between the SUSY scale and 1 GeV\cite{long}. All of these have been investigated
thoroughly in \cite{Nano,Muraya}.
The $\tilde{c}$ and $\tilde{t}$-exchange amplitudes can be constructive or destructive with each other depending on $\phi_c$ and $\phi_t$.
The ratio of $\tilde{t}$ and $\tilde{c}$-contribution can be defined by\cite{Muraya}
\begin{equation}
y^{tK_\mu}=\frac{Y^{32} A_S(t,u,s)e^{i\phi_t}  V_{ts} V_{td} (f(t,\,\mu) + f(t,\,d'))}{ Y^{22} A_S(c,u,s)e^{i\phi_c} V_{cs} V_{cd} (f(c,\,\mu) + f(c,\,d'))}.
\end{equation}
We also have
\begin{equation}
y^{tK_e}=\frac{Y^{31}Y^{22}}{Y^{21}Y^{32}}y^{tK_\mu}
\end{equation}
for $p\to K^++\bar\nu_e$. This process is suppressed by the smallness
of the first generation Yukawa couplings in the MSSM.

The couplings $Y^{ij}$'s are fixed by GUT-scale masses and light fermion Yukawa couplings.
With fixed GUT-scale masses, we have the parameter freedom to set $Y^{12}=0$, $Y^{22}=0$, and $Y^{32}=0$
to suppress $p\to K^++\bar\nu_\mu$ by tuning $h^{\prime}$'s. Inversely, we can get 45-Higgs Yukawa couplings up to GUT-scale masses
from proton decay bounds. If $p\to K^++\bar\nu_\mu$ is suppressed, the
modes $p\to \text{mesons}+\bar\nu_\mu$ and $p\to
\text{mesons}+\mu^+$ are all suppressed.

\section{\label{sec:gutc}Constraints in Small Mixing
Limit of 5 and 45-plets}

This model has two pairs of colored Higgs multiplets which can induce proton decay. The contributions from different Higgs multiplets can be destructive, so that it may predict long enough proton lifetimes. This destruction can be exhibited in a very simple way when the mixing between different Higgs multiplets can be treated as perturbation. When the mixing is negligible, we can take the masses of the color Higgs to be degenerate as the leading order approximation. The Yukawa couplings $Y^{ij}$ can be expanded through mixing angle $\theta_T$ and the mass difference of the colored Higgs sectors beyond the leading order. They are related to $\rho$, $\kappa_{1 ,2}$ and $t$, so we can expand dimension-five operator couplings through these parameters. The couplings $h^{(\prime)i}$ and $f^{(\prime)i}$ are restricted by Eq.(\ref{fyuka}).

\subsection{Leading Order Cancellation}
In this subsection, we pick out the most relevant contributions and consider their counteractions. 
When the mixings between 5 and 45-plets are negligible, we have the leading terms of $Y^{ij}$'s as follows
\begin{eqnarray}
Y^{ij}_{(0)}&=&\frac{h^if^j}{2M_{H_c}}+\frac{2h^{\prime i}f^{\prime j}}{M_{H_c'}}.
%X^{ij}_{(0)}&=&\frac{h^if^j}{M_{H_c}}+\frac{4h^{\prime i}f^{\prime j}}{M_{H_c'}}.
\end{eqnarray}

If we further assume \begin{equation}\label{yadhoc}
h^if^j+4h^{\prime i}f^{\prime i}=0,
 \end{equation}
 and \begin{equation}\label{madhoc}M=M_{H_c}=M_{H_c'},\end{equation} $Y^{ij}_0$ does not contributes to the nucleon decays. $h^{(\prime i)}$ are specially chosen to let the contributions from different colored Higgs sectors cancel. In such way, we illustrate the destructive effect. Then the next to leading order terms dominate. The $\sin^2\theta_T$ terms do not contribute to the next to leading order terms. The $\sin\theta_T\cos\theta_T$ terms vanish when $M_{H_c}=M_{H_c'}$. The next to leading order terms only come from the colored Higgs mass splitting and can be written as
\begin{eqnarray}\label{yl1}
Y^{ij}_{(1)}&=&\frac{h^if^j}{2M^2}(M_{H_c'}- M_{H_c}).
%X^{ij}_{(1)}&=&\frac{h^if^j}{M^2}(M_{H_c'}- M_{H_c}).
\end{eqnarray}
%In this case, the leading contributions only come from the extra LLLL-type operators. The decay width is proportional to the inverse square of $H^a_{b\alpha}$'s mass
%\begin{equation}\label{ga1}
%\Gamma\propto \frac{1}{M^2_{H^{a}_{b\alpha}}}.
%\end{equation}

%Now we study the destructive effects among all the colored Higgs contributions.
%Still we are working under the assumption that the mixing between 5 and
%45-plets is negligible. This condition results in the simple forms of
%dimension-five operator couplings. The leading terms of
%Eq.(\ref{d5coupling}) can be written as
%\begin{equation}\label{yukafix}
%y^{ij}_{(0)}=\frac{h^if^i}{2M_{H_c}}+\frac{2h^{\prime i}f^{\prime
%    i}}{M_{H_c'}}-\frac{h^{\prime i}f^{\prime i}}{M_{H^a_{b\alpha}}}.
%\end{equation}
%If we set the Higgs masses  satisfying
%\begin{equation}\label{madhoc}
%M=M_{H_c}=M_{H_c'}=M_{H^a_{b\alpha}},
%\end{equation}and the Yukawa couplings
%\begin{equation}\label{yadhoc}
%h^if^j+2h^{\prime i}f^{\prime j}=0,
%  \end{equation}
%we have $y^{ij}_{(0)}=0$.  Comparing with Eq.(\ref{mass}), we find that
%the second equals sign of Eq.(\ref{madhoc}) requires $|t|\ll |s|$ and the last equals sign requires $|\rho V|\ll \mu_2$.
%The above conditions ensure that there is no dimension-five operator mediated proton decay in the leading order of $y^{ij}$.

\subsection{Beyond the Leading Order}
Even the leading order of $Y^{ij}_0$ vanishes, the non-vanishing $\kappa$ and the colored Higgs multiplets mass splitting can still contribute to nucleon decay.
%The next to leading order terms of $Y^{ij}$ are
%\begin{eqnarray}\label{yl1}
%Y^{ij}_{(1)}&=-&\frac{\delta M_{H_c}}{M^2}\frac{h^if^j}{2}-\frac{2\delta
%  M_{H_c'}}{M^2}h^{\prime i}f^{\prime j}\nonumber\\%+\frac{\delta  M_{H^a_{b\alpha}}}{M^2}h^{\prime i}f^{\prime j},\nonumber\\
%&=&\frac{h^if^j}{2M^2}\left(\delta M_{H_c}+2\delta M_{H^a_{b\alpha}}-4\delta
%  M_{H_c'}\right).\nonumber\\
%&=&-\frac{h^if^j}{2M^2}\left(M_{H_c}-2 M_{H_c'}\right).
%\end{eqnarray}
%where $\delta M_{H_{H_c}}=M_{H_{H_c}}-M$ and the other two $\delta$'s are defined in a similar way.
In order to investigate the validity of $Y_{(1)}$, we  need the next to the next to leading order terms
\begin{eqnarray}\label{2rd}
Y^{ij}_{(2)}&=&\frac{ M_{H_c'}- M_{H_c}}{M^2}\frac{\sqrt{2}}{4}\left(h^if^{\prime j}+8h^{\prime i}f^j\right)\sin\theta_T.%-\frac{h^if^j}{2M}\sin^2\theta_T.%-\frac{(\delta M_{H^a_{b\alpha}})^2}{M^3}\frac{h^if^j}{2},
\end{eqnarray}
For small $|\kappa_{1(2)}|\ll|t|$, we have
\begin{equation}
\sin\theta_T\approx \frac{\kappa_1+\kappa_2}{\sqrt{2}t},
\end{equation}
and
\begin{eqnarray}\label{hcdif}
M_{H_c}-M_{H_c'}&\approx& |st|/4.%\nonumber\\
%\delta M_{H^a_{b\alpha}}&\approx& 2\rho V.
\end{eqnarray}
When $|t|\ll1$, $|\kappa_{1(2)}|\ll |t|$,  $Y^{ij}_{(2)}$ is negligible and $Y^{ij}_{(1)}$ is a good approximation to $Y^{ij}$.
In this special circumstance, we have the nucleon decay width
\begin{equation}\label{propto}
\Gamma\propto\frac{(M_{H_c}- M_{H_c'})^2}{M^4}.
\end{equation}
It is proportional to the squared mass splitting.

We can also simplify $y^{tK}$ when $Y^{32}_{(0)}=0$ and
$Y^{22}_{(0)}=0$ or $Y^{31}_{(0)}=0$ and $Y^{21}_{(0)}=0$ . The most
relevant term is
\begin{eqnarray}
|y^{tK}|=|y^{tK_\mu}|=|y^{tK_e}|=|\frac{m_tV_{td}V_{ts}}{m_cV_{cd}V_{cs}}|\approx 0.18,
\end{eqnarray}
with common triangle diagram factors and short range renormalization factor.

It is important to emphasize that we can only suppress some specific decay
modes in this way. When one mode is suppressed others may be
enhanced. For example, if we choose $Y^{32}=0$ and $Y^{22}=0$ to
suppress $p\to K^++\nu_\mu$,  $p\to \pi^++\nu_e$ will be enhanced
for the consequential larger $Y^{31}$ and  $Y^{21}$.

When the mixing effects between 5 and 45-plets dominate,
the mass difference between  ${H_c}$ and ${H_c'}$ will not be small.
We can not expand the dimension-five operator couplings
in the colored Higgs mass splitting. The most dominant part of proton
decay widths can not be picked out in a simple way.

Under the conditions of this section, the heavy thresholds effects in Eqs.(\ref{gut1}) and (\ref{gut2}) are simplified. When $|\rho V|\ll \mu_2$, only $M_{H_c}$, $M_{H_c'}$, $M_{H_c''}$, and $M_{H_D}$ contribute to Eq.(\ref{gut1}) and only $M_{V}$ and $M_{\Sigma}$ contribute to Eq.(\ref{gut2}). All the other GUT-scale Higgses contributions offset each other, because their masses all approximately equal to $\mu_2$ and appear in logarithmic functions with different signs. The masses contribute to nucleon decay could be determined from Eqs.(\ref{rcond0}) and (\ref{rcond1}). Together with the conditions $|t|\ll s$ and $\kappa$'s$\ll\mu_2$, numerical study show that all the GUT-scale masses are around $1\times 10^{15}$GeV. The limits that we taken here are fully compatible with the constraints of Sec. \ref{sec:gutm}.

\section{\label{sec:num}Numerical Constraints on Proton Decay}

In this section, we present some numerical results about the destructive effect. We illustrate it in the small mixing limit at first, and then in a relaxed parameter space. We choose some specific proton decay modes for simplicity. The most relevant modes $p\to
K^++\nu$ and the modes without dependence on $e^{i\phi}$ are our
concentration.  At first we analyze these models in the small mixing limit of 5
and 45-plet following the general discussion of
Sec. \ref{sec:gutc}. Then we use the predictions from
previous sections to get the possible longest proton lifetimes of this model.

We take all the sparticle masses at 1TeV except for $m_{\tilde d}=m_{\tilde \mu}=10$TeV\cite{Muraya3}. We neglect squark and slepton mixing for simplicity. Because most of the mass insertion parameters are small \cite{Ko2008}, we will not loose the major physics. Present limit on chargino mass is $m_{\tilde\chi^{\pm}}>94$GeV\cite{PDG}. In this work we take wino mass $M_2=100$GeV.
The present experimental limits on partial proton lifetimes  at 90\% C.L.\cite{PDG} are
\begin{eqnarray}\label{exp}
\tau(p\to \pi^0+e^+)&>&1.6\times10^{33}\text{yrs}\nonumber \\
\tau(p\to K^++\bar\nu)&>&6.7\times10^{32}\text{yrs}\nonumber \\
\tau(p\to \pi^0+\mu^+)&>&4.73\times10^{32}\text{yrs}
\end{eqnarray}

We also take $M_{GUT}\sim2\times 10^{16}$GeV and fix $M_V=4.96\times10^{16}$GeV and $M_\Sigma=2.5\times10^{15}$GeV. The fermion masses are taken as input through Eqs.(\ref{fmass}, \ref{fyuka}). The CKM matrix parameters $V^*_{ij}$ are taken from Ref.\cite{PDG}. The Yukawa couplings $f^i$ and $f^{\prime i}$ are fixed up to the doublets' mixing angle $\theta_D$ and $\tan\beta$. In the small mixing limit, they are fixed by $\tan\beta$ for $\theta_D\sim 0$. The up-type quark masses, $\theta_D$, and $\tan\beta$ can not fix the up-type Yukawa couplings. By tuning $h^i$ and $h^{\prime i}$, we can get destructive effects among colored Higgs contributions.

After fixing all the Yukawa couplings, we can use the chiral Lagrangian technique\cite{Claudson,Chadha} to transform quark level Lagrangian in Eq.(\ref{fl}) to the hadronic level.  This is accomplished through the matrix elements
\begin{equation}
\langle K^+|(u,d)_L s_L|p\rangle=\frac{\beta}{f}\left(1+\left(\frac{D}{3}+F\right)\frac{m_N}{m_B}\right),\label{had1}
\end{equation}
\begin{equation}
\langle K^+|(u,s)_L d_L|p\rangle=\frac{\beta}{f}\frac{2D}{3}\frac{m_N}{m_B},\label{had2}
\end{equation}
which is obtained in the limit $m_{u,d,s}\ll m_{N,B}$. All the parameters can be found in \cite{Muraya,aoki}. We take them as following \cite{aoki}
\begin{eqnarray}
\beta=0.0118\text{GeV}^3,~~ D=0.8,~~ F=0.47,~~f=0.131\text{GeV},~~
 m_N=0.94\text{GeV},~~ m_B=1.15.\text{GeV}.
\end{eqnarray}

In Table \ref{tab:proton}, we list the
general form of dimension-five operators mediated partial proton lifetimes. These formal equations do not depend on the assumption of parameters. We only have to determine the values of $y^{ij}$ and $y^{tK}$ to get the proton decay rates.

\begin{table}[h]
\caption{\label{tab:proton} The dominant proton decay modes and the
  modes independent of $e^{i\phi}$ or $y^{tK}$ without  ad hoc assumption. We assume the dimension-five operator contributions dominate.}
%\begin{ruledtabular}
\begin{tabular}{rrl}
\hline\hline
Decay mode& General&Lifetime\\
\hline
$\tau(p\to K^++\bar\nu_\mu)$&$1.0\times10^{35}$& \\
$\tau(p\to K^++\bar\nu_e)$&$5.3\times10^{33}\left|\frac{\displaystyle Y^{22}(1+y^{tK_\mu})}{\displaystyle Y^{21}(1+y^{tK_e})}\right|^2$&\raisebox{2.3ex}[0pt]{$\times\left|\frac{\displaystyle 10^{-25}\text{GeV}}{\displaystyle Y^{22}(1+y^{tK_\mu})}\frac{\displaystyle \text{TeV}^{-1}}{\displaystyle A_s(f(c,\,\mu)+ f(c,\,d'))}\right|^{2}$yrs} \\
\hline
$\tau(p\to \pi^0+\mu^+)$&$3.2\times10^{36}$&\\
$\tau(p\to \pi^0+e^+)$&$7.8\times10^{34}\left|\frac{\displaystyle Y^{12}}{\displaystyle Y^{11}}\right|^2$&\raisebox{2.3ex}[0pt]{$\times\left|\frac{\displaystyle 10^{-25}\text{GeV}}{\displaystyle Y^{12}}\frac{\displaystyle \text{TeV}^{-1}}{\displaystyle A_s(f(u,\,d')+ f(d',\,\nu))}\right|^{2}$yrs} \\
\hline\hline
\end{tabular}
%\end{ruledtabular}
\end{table}

\subsection{Proton Lifetime in Small Mixing Limit}

Now we numerically study the destructive effects among all colored Higgs in the small mixing limit. Eq.(\ref{yadhoc}) ensures the most relevant contribution comes from Eq.(\ref{yl1}).
For the decay modes we concern, their ambiguities reside in the parameters $Y^{22}$, $Y^{12}$, ${Y^{22}}/{Y^{21}}$, ${Y^{12}}/{Y^{11}}$, and $y^{tK}$.
$y^{tK}$ contains an unknown phase $e^{i(\phi_t-\phi_c)}$. This phase is a free parameter at present and may bring cancellation between c and t quark contributions.
It follows $y^{tK_\mu}\approx y^{tK_e}$ under the conditions of the small
mixing limit.
When $Y^{22}_{(0)}=Y^{12}_{(0)}=0$, we have
\begin{equation}
\frac{Y^{12}}{Y^{11}}\approx\frac{Y^{22}}{Y^{21}}\approx \frac{f^2f^{\prime 2}}{f^{\prime 2}f^1-f^{2}f^{\prime 1}}\frac{M_{H_c'}-  M_{H_c}}{M},
\end{equation}
which can be obtained directly from Eq.(\ref{yl1}). %Furthermore we can conclude that the fractional part of the Yukawa couplings only depend on the down-type quark and lepton masses
Numericall, we have
\begin{equation}
\frac{f^2f^{\prime 2}}{f^{\prime 2}f^1-f^{2}f^{\prime 1}}=(-4.9\sim7.1).
\end{equation}
by applying Eq.(\ref{fyuka}).
The mass parameters are from Ref.\cite{PDG}.
Now we only have to consider $Y^{22}$ and $Y^{12}$. We follow the discussion in Sec. \ref{sec:gutc}.
With $Y^{22}_{(0)}=0$ and Eqs.(\ref{yadhoc}) and (\ref{yl1}), we have
\begin{eqnarray}
Y^{22}_{(1)}&=&-\frac{\pi\alpha_2m_c(3m_s+m_\mu)}{m_W^2\sin2\beta(\cos^2\theta_D+\frac{3m_s+m_\mu}{32(m_\mu-m_s)}\sin^2\theta_D)}\nonumber\\
&&\times\frac{1}{2M^2}\left(M_{H_c'}- M_{H_c}\right)\nonumber\\
&\approx&\frac{\pi\alpha_2m_c(3m_s+m_\mu)}{m_W^2\sin2\beta}\frac{1}{2M^2}\left(M_{H_c}- M_{H_c'}\right).
\end{eqnarray}
$Y^{12}_{(1)}$ can be obtained by replacing $m_c$ with $m_u$. The numerical study gives
\begin{equation}
Y^{22}_{(1)}\approx\frac{(7\sim10)\times10^{-6}}{\sin2\beta}\frac{M_{H_c}- M_{H_c'}}{2M^2}.
\end{equation}
The range comes from the uncertainties in quark masses.
When $Y^{21}_{(0)}=0$, we have a similar analysis.
Then the last uncertainty comes from the colored Higgs masses.  We
take $\tan\theta$ as small as possible. We further take $|t|<|s|$. In
this way, we approach the condition of Sec.\ref{sec:gutc}. At this time, the GUT-scale masses are all fix at around $10^{15}$GeV, which are listed in Table \ref{tab:sumw}.  We have the possible longest
proton lifetimes in Table \ref{tab:limit}.
When one of the modes $p\to K^++\nu_\mu$ or $p\to K^++\nu_e$ reaches its maximum, the other one gets their minimum.
This can be explained by the fact that the condition $Y^{22}_{(0)}=0$ enhances
$Y^{21}_{(0)}$ and $Y^{21}_{(0)}=0$ enhances
$Y^{22}_{(0)}$, which have been explained in Sec. \ref{sec:gutc}. The same explanation is also applicable for $p\to \pi^0+\mu^+$ and $p\to \pi^0+e^+$.

\begin{table}[h]
\caption{\label{tab:sumw} The GUT-scales masses in small mixing limit. }
%\begin{ruledtabular}
\begin{tabular}{lcr}
\hline
\hline
Higgs&Representation&Mass(GeV)\\
\hline
${H_c}\,\bar H_c$ &$(1,3)_{-\frac{1}{3}}\,(1,\bar3)_\frac{1}{3}$&$1.22\times10^{15}$\\
${H_c'}\,{\bar H_c'}$&$(1,3)_{-\frac{1}{3}}\,(1,\bar3)_\frac{1}{3}$&$1.219\times10^{15}$\\
${H^a_{b\alpha}}\,H_a^{b\alpha}$&$(3,\bar3)_{\frac{1}{3}}\,(3,3)_{-\frac{1}{3}}$&$1.07\times10^{15}$\\
${H_c''\, \bar H_c''}$&$(1,3)_{-\frac{4}{3}}\,(1,\bar3)_{\frac{4}{3}}$&$1.23\times10^{15}$\\
${H^\alpha_{\beta a}}\, H_\alpha^{\beta a}$&$(2,8)_{-\frac{1}{2}}\,(2,8)_{\frac{1}{2}}$&$1.23\times10^{15}$\\
${H^{a\beta}}\,H_{a\beta}$&$(2,3)_{\frac{7}{6}}\,(2,\bar 3)_{-\frac{7}{6}}$&$1.17\times10^{15}$\\
${H^{\alpha\beta}_{(s)}}\,H_{\alpha\beta}^{(s)}$&$(1,6)_{\frac{1}{3}}\,(1,\bar 6)_{-\frac{1}{3}}$&$1.33\times10^{15}$\\
${H_u'}\,H_d'$&$(2,1)_{\frac{1}{2}}\,(2,1)_{-\frac{1}{2}}$&$1.01\times10^{15}$\\
\hline
\hline
\end{tabular}
\end{table}

\begin{table}[h]
\caption{\label{tab:limit} The longest partial proton lifetimes in the small mixing limit with $\tan\beta=2$.
The first four ones are gotten with $Y^{22}_{(0)}=0$. The last four ones are gotten  with $Y^{21}_{(0)}=0$.}
%\begin{ruledtabular}
\begin{tabular}{cc}
\hline
\hline
Decay mode &  Lifetime(yrs)\\
\hline
$\tau(p\to K^++\bar\nu_\mu)$&$<3.7\times10^{36}$\\
$\tau(p\to K^++\bar\nu_e)$&$<1.4\times10^{29}$\\
$\tau(p\to \pi^0+\mu^+)$&$<3.3\times10^{44}$\\
$\tau(p\to \pi^0+e^+)$&$<1.9\times10^{37}$\\
\hline
$\tau(p\to K^++\bar\nu_\mu)$&$<1.3\times10^{31}$\\
$\tau(p\to K^++\bar\nu_e)$&$<7.9\times10^{38}$\\
$\tau(p\to \pi^0+\mu^+)$&$<1.4\times10^{38}$\\
$\tau(p\to \pi^0+e^+)$&$<1.2\times10^{40}$\\
\hline
\hline
\end{tabular}
%\end{ruledtabular}
\end{table}

\subsection{Proton Lifetime beyond Small Mixing Limit}
Although this model does not satisfy the experiments in the small mixing limit, we can analyze larger parameter space with the small mixing limit as our starting point. In the small mixing limit, we can not realize the smallness of $Y^{22}$ and $Y^{21}$ at the same time. They are related to the two most dominant modes $p\to
K^++\nu_\mu$ and $p\to K^++\nu_e$. If we relax the parameter space a little, we can find some specific choice of $h^i/h^{\prime i}$'s to ensure the long enough proton lifetime. This could be verified by the following analysis.

We begin with Eq.(\ref{madhoc}) which gives $H_c$ and $H_c'$ a common mass $M$. At this time all the GUT-scale Higgs sectors are around $M$. When Eq.(\ref{madhoc}) is exactly satisfied, together with Eq.(\ref{yadhoc}) we have $Y^{ij}=0$. This result directly comes from the form of $Y^{ij}$ and does not depend on any other assumptions. But it is impossible to let all $Y^{ij}$ equal to zero at the same time, e.g.  $Y^{22}$ will be large enough to break experimental bound when we set $Y^{21}=0$ with $h^2/h^{\prime 2}=-4f^{\prime 1}/f^1$. We can not even find proper  $h^2$ and $h^{\prime 2}$ to let $Y^{21}$ and $Y^{22}$ both small enough. But if we are able to set $H_c$, $H_c'$, and ${H_{b\alpha}^a}$ at a higher scale, this problem can be solved.

Since we want to relax the small mixing limit in a controlled way, we change as few parameters as possible and take the process little by little. We set $\mu_2$ to be larger. Then the parameters $s$ and $|t|$ becomes larger. Hence, the mass of ${H_c}$ and ${H_c'}$ becomes heavier and $H_c$ is heavier than $H_c'$. $H_D'$ becomes heavier too. The gauge coupling unification constraint Eq.(\ref{rcond1}) of Sec. \ref{sec:gutm} sets constraints on the masses of  ${H_c}$, ${H_c'}$, and $H_D'$. At this time, Eq.(\ref{mcond}) will make $\kappa_{1}$ or $\kappa_2$ larger if we do not change $\mu_1$ smaller. In order to keep mass difference between ${H_c}$ and ${H_c'}$ not too large, we need to keep $|t|\ll s$ and hence $\mu_1$ should not becomes smaller. Accordingly, $\kappa_1$ or $\kappa_2$ becomes larger and we break the small mixing condition. In a word, we have all the three colored Higgs sectors becomes heavier at the expanse of the small mixing limit.

%we vary $M_{H_c}$,$M_{H_c'}$, and $M_{H_{b\alpha}^a}$ from each other a little. the unification constraints of Sec. \ref{sec:gutm} can push these Higgs sectors to a higher scale. This could be easily seen from Eq.(\ref{rcond1}). In the small mixing limit, the Higgs sectors $H_c$, ${H_c'}$, and ${H_D'}$ are all below $3.6\times10^{15}$GeV and  $M_{H_{b\alpha}^a}<3.6\times10^{15}$GeV consequently. If we take $H_c$, ${H_c'}$ at a little scale,  $M_{H_D'}$ can become larger enough. Hence, $H_c$ and ${H_c'}$ is pushed to a higher scale. Since we start with the small mixing limit

Now, it is possible to realize the smallness of $Y^{22}$ and $Y^{21}$ at the same time under the constraints of Sec. \ref{sec:gutm}. %These two conditions conflict with each other. They are related to the two most dominant modes $p\to K^++\nu_\mu$ and $p\to K^++\nu_e$. However, we find that the proton lifetime of this model can agree with the experimental results with specific choice of parameters.
The real destruction effects in $Y^{ij}$ can still be achieved with specific choice of $h^i/h^{\prime i}$'s.  % to ensure the long enough proton lifetime.
The possible longest partial lifetimes are listed in Table \ref{tab:long} and the corresponding GUT-masses are in Table \ref{tab:sumnw}. These GUT-masses are gotten by the restriction from Sec.\ref{sec:gutm} and by relaxing the small mixing limit. The proton lifetimes are in accord with the data in Eq.(\ref{exp}).

\begin{table}[h]
\caption{\label{tab:long} The longest partial proton lifetimes with $\tan\beta=2$.
The lower limit of $p\to \pi^0+e^+$ comes from X and Y gauge boson. The upper limit comes from dimension-five operator mediation.}
%\begin{ruledtabular}
\begin{tabular}{cc}
\hline
\hline
Decay mode&Lifetime\\
\hline
$\tau(p\to K^++\bar\nu)$&$<1.6\times10^{33}$yrs\\
$\tau(p\to \pi^0+\mu^+)$&$<1.7\times10^{39}$yrs\\
$\tau(p\to \pi^0+e^+)$&$<5.9\times10^{46}$yrs(d=5)\\
$\tau(p\to \pi^0+e^+)$&$>5.7\times10^{34}$yrs(d=6)\\
\hline
\hline
\end{tabular}
%\end{ruledtabular}
\end{table}

\begin{table}[h]
\caption{\label{tab:sumnw} The GUT-scales masses which ensure longest proton lifetime without small mixing limit. These masses are in accord with the gauge coupling unification constraints of Sec. \ref{sec:gutm}.}
%\begin{ruledtabular}
\begin{tabular}{lcr}
\hline
\hline
Higgs&Representation&Mass(GeV)\\
\hline
${H_c}\,\bar H_c$ &$(1,3)_{-\frac{1}{3}}\,(1,\bar3)_\frac{1}{3}$&$1.48\times10^{16}$\\
${H_c'}\,{\bar H_c'}$&$(1,3)_{-\frac{1}{3}}\,(1,\bar3)_\frac{1}{3}$&$3.07\times10^{15}$\\
${H^a_{b\alpha}}\,H_a^{b\alpha}$&$(3,\bar3)_{\frac{1}{3}}\,(3,3)_{-\frac{1}{3}}$&$1.03\times10^{16}$\\
${H_c''\, \bar H_c''}$&$(1,3)_{-\frac{4}{3}}\,(1,\bar3)_{\frac{4}{3}}$&$9.56\times10^{15}$\\
${H^\alpha_{\beta a}}\, H_\alpha^{\beta a}$&$(2,8)_{-\frac{1}{2}}\,(2,8)_{\frac{1}{2}}$&$9.56\times10^{15}$\\
${H^{a\beta}}\,H_{a\beta}$&$(2,3)_{\frac{7}{6}}\,(2,\bar 3)_{-\frac{7}{6}}$&$9.87\times10^{15}$\\
${H^{\alpha\beta}_{(s)}}\,H_{\alpha\beta}^{(s)}$&$(1,6)_{\frac{1}{3}}\,(1,\bar 6)_{-\frac{1}{3}}$&$9.18\times10^{15}$\\
${H_u'}\,H_d'$&$(2,1)_{\frac{1}{2}}\,(2,1)_{-\frac{1}{2}}$&$1.24\times10^{16}$\\
\hline
\hline
\end{tabular}
\end{table}

\section{Summary}
In this work we analyze the SUSY SU(5) GUT model with $5+\bar 5$ and $45+\overline{45}$ Higgs multiplets. We give the most general renormalizable superpotential for the Higgs fields and regain the light fermion mass formulae.
%We also find an explanation for the "$\mu$ problem" through the see-saw mechanism.
We set constraints on GUT-scale masses through gauge coupling unification. We analytically study the possibility that dimension-five operators from different Higgs sectors are destructive, so we can have long enough proton lifetime. When the mixings between 5 and 45-plet Higgs fields can be neglected and the masses of the Higgs multiplets are  degenerate, the proton decay width can be studied in a quite simple manner. Although this limit does not satisfy the experimental bounds, we can relax this limit and still maintain the destructive effects.  In a general parameter space, we find that the proton lifetime can be in agreement with the present experimental bounds.

This work was supported in part by the National Natural Science
Foundation of China (NSFC) under Grant No. 10435040.

\bibliographystyle{JHEP}

\bibliography{su5}

\providecommand{\href}[2]{#2}\begingroup\raggedright\begin{thebibliography}{10}

\bibitem{luo}
P.~Langacker and M.~Luo, {\it Implications of precision electroweak experiments
  for $m_t$, $\rho{}$, $sin^2\theta{}_w$, and grand unification},  {\em Phys.
  Rev. D} {\bf 44} (1991) 817.

\bibitem{Ellis1991131}
J.~Ellis, S.~Kelley, and D.~V. Nanopoulos, {\it Probing the desert using gauge
  coupling unification},  {\em Phys.\ Lett.\ B} {\bf 260} (1991) 131.

\bibitem{Georgi:1979df}
H.~Georgi and C.~Jarlskog, {\it A new lepton - quark mass relation in a unified
  theory},  {\em Phys. Lett. B} {\bf 86} (1979) 297.

\bibitem{sakai}
N.~Sakai and T.~Yanagida, {\it Proton decay in a class of supersymmetric grand
  unified models},  {\em Nucl.\ Phys.\ B} {\bf 197} (1982) 533.

\bibitem{Weinberg}
S.~Weinberg, {\it Supersymmetry at ordinary energies. masses and conservation
  laws},  {\em Phys.\ Rev.\ D} {\bf 26} (1982) 287.

\bibitem{susygj}
S.~Dimopoulos, L.~J. Hall, and S.~Raby, {\it Predictive framework for fermion
  masses in supersymmetric theories},  {\em Phys. Rev. Lett.} {\bf 68} (1992)
  1984.

\bibitem{Raby:1992vk}
S.~Raby, {\it {Fermion masses in SUSY GUTs}},  {\em Presented at 27th
  Rencontres de Moriond on Electroweak Interactions and Unified Theories, Les
  Arcs, France, Mar} (1992) 15.

\bibitem{Muraya3}
H.~Murayama and A.~Pierce, {\it Not even decoupling can save the minimal
  supersymmetric su(5) model},  {\em Phys. Rev. D} {\bf 65} (2002) 055009.

\bibitem{bajc}
B.~Bajc, P.~Fileviez~P\'erez, and G.~Senjanovi\'c, {\it Proton decay in minimal
  supersymmetric su(5)},  {\em Phys. Rev. D} {\bf 66} (2002) 075005.

\bibitem{PhysRevD.77.015015}
P.~Nath and R.~M. Syed, {\it Suppression of higgsino mediated proton decay by
  cancellations in grand unified theories and strings},  {\em Phys. Rev. D}
  {\bf 77} (2008) 015015.

\bibitem{nonmini}
P.~Kalyniak and J.~N. Ng, {\it Symmetry-breaking patterns in su(5) with
  nonminimal higgs fields},  {\em Phys. Rev. D} {\bf 26} (1982) 890.

\bibitem{nonm3}
P.~Eckert, J.~M. Gerard, H.~Ruegg, and T.~Schuecker, {\it Minimization of the
  su(5) invariant scalar potential for the 45-dimensional representation},
  {\em Phys. Lett. B} {\bf 125} (1983) 385.

\bibitem{adj}
P.~Fileviez~P\'erez, H.~Iminniyaz, and G.~Rodrigo, {\it Proton stability, dark
  matter, and light color octet scalars in adjoint su(5) unification},  {\em
  Phys. Rev. D} {\bf 78} (2008) 015013.

\bibitem{adj2}
P.~Fileviez~P\'erez, {\it Supersymmetric adjoint $su(5)$},  {\em Phys. Rev. D}
  {\bf 76} (2007) 071701.

\bibitem{Muraya}
J.~Hisano, H.~Murayama, and T.~Yanagida, {\it Nucleon decay in the minimal
  supersymmetric su(5) grand unification},  {\em Nucl.\ Phys.\ B} {\bf 402}
  (1993) 46.

\bibitem{Muraya2}
J.~Hisano, H.~Murayama, and T.~Yanagida, {\it Probing the
  grand-unification-scale mass spectrum through precision measurements on the
  weak-scale parameters},  {\em Phys. Rev. Lett.} {\bf 69} (1992) 1014.

\bibitem{Barbieri}
R.~Barbieri and L.~J. Hall, {\it Grand unification and the supersymmetric
  threshold},  {\em Phys. Rev. Lett.} {\bf 68} (1992) 752.

\bibitem{Ellis}
J.~Ellis, S.~Kelley, and D.~V. Nanopoulos, {\it Constraints from gauge coupling
  unification on the scale of supersymmetry breaking},  {\em Phys.\ Lett.\ B}
  {\bf 287} (1992) 95.

\bibitem{Nath}
P.~Nath and P.~Fileviez~P\'erez, {\it Proton stability in grand unified
  theories, in strings and in branes},  {\em Phys. Rep.} {\bf 441} (2007) 191.

\bibitem{Georgi}
H.Georgi, {\it in particles and fields-1974},  {\em Proceedings of the
  Williamsburg Meeting of the Division of Particles and Fields of the American
  Physical Society, edited by C. E. Carlson} (AIP, New York, 1975) 575.

\bibitem{Goto}
T.~Goto and T.~Nihei, {\it Effect of an rrrr dimension 5 operator on proton
  decay in the minimal su(5) sugra gut model},  {\em Phys.\ Rev.\ D} {\bf 59}
  (1999) 115009.

\bibitem{Arno}
P.~Nath and R.~Arnowitt, {\it Limits on photino and squark masses from proton
  lifetime in supergravity models},  {\em Phys.\ Rev.\ D} {\bf 38} (1988) 1479.

\bibitem{long}
T.~Nihei and J.~Arafune, {\it The two loop long range effect on the proton
  decay effective lagrangian},  {\em Prog.\ Theor.\ Phys.} {\bf 93} (1995) 665.

\bibitem{Nano}
J.~Ellis, D.~V. Nanopoulos, and S.~Rudaz, {\it A phenomenological comparison of
  conventional and supersymmetric guts},  {\em Nucl.\ Phys.\ B} {\bf 202}
  (1982) 43.

\bibitem{Ko2008}
P.~Ko, J.~hyeon Park, and M.~Yamaguchi, {\it Sflavor mixing map viewed from a
  high scale in supersymmetric su(5)},  {\em JHEP} {\bf 11} (2008) 051.

\bibitem{PDG}
{\bf Particle Data Group} Collaboration, C.~Amsler {\em et.~al.}, {\it {Review
  of particle physics}},  {\em Phys. Lett. B} {\bf 667} (2008.) 1.

\bibitem{Claudson}
M.~Claudson, M.~B. Wise, and L.~J. Hall, {\it Chiral lagrangian for deep mine
  physics},  {\em Nucl.\ Phys.\ B} {\bf 195} (1982) 297.

\bibitem{Chadha}
S.~Chadha and M.~Daniels, {\it Chiral lagrangian calculation of nucleon decay
  modes induced by d=5 supersymmetric operators},  {\em Nucl.\ Phys.\ B} {\bf
  229} (1983) 105.

\bibitem{aoki}
Y.~Aoki, C.~Dawson, J.~Noaki, and A.~Soni, {\it Proton decay matrix elements
  with domain-wall fermions},  {\em Phys.\ Rev.\ D} {\bf 75} (2007) 014507.

\end{thebibliography}\endgroup
\end{document}